\documentclass[preprint,showpacs,preprintnumbers,amsmath,amssymb]{revtex4}

\input{psfig}
\begin{document}
\preprint{APS/}
\title{  Adsorption vs. folding of a  Hydrophobic Chain Protein Model near   an Attractive Surface}
\author{  Handan Arkin and Hakan Alaboz }
\email{holgar@eng.ankara.edu.tr}
\affiliation{ Ankara University, Faculty of Engineering, Department of Physics Engineering
               Tando\u{g}an, Ankara, Turkey  }
\begin{abstract}

The folding vs. adsorption behaviour of a coarse-grained off-lattice  protein model near  an attractive surface  is presented within the frame  of a  Multicanonical Monte Carlo simulations. In the polymer-surface model, the Lennard-Jones
potential is assumed as an interaction potential between the effective monomers and the attractive surface. Thermodynamic properties and some structural parameters for the minimum energy conformations are  calculated for comparison of the folding and adsorption cases.

\end{abstract}

\maketitle

\section{Introduction}

Adsorption of polymers on flat surfaces and geometries is a longstanding challenge that draw attention
of researcher from different areas and  has many applications varying from adhesion~\cite{Steiner},
pattern recognition~\cite{nakata, bogner}, biomedical implant motification to protein ligand binding and docking~\cite{Service,balog, ike, gupta}. The arrengement of polymers on the surface attracts attention because of it is an onset of electronic circuits in nanotechnological device invention~\cite{nakata, Bachmann1, Bachmann2}. Therefore, too much theoretical and experimental study focus on how the binding affinity of proteins and polymers affect the adsorption phenomena. In these content, there are many interesting and important problems, which are concerned with the general aspects of the questions how the proteins fold or adsorb on the surface~\cite{Gokoglu}. If we understand the folding vs. adsorption mechanism of a polymer or protein on a flat surface, then it would be possible to engineer desirable biologically active surfaces with specific properties~\cite{PRE73R}.

Surprisingly, the self-assembly of proteins and polymers near a surface is not studied in detail
because the length and time scale of relevance largely conflict 
with experimental techniques and full atom computer simulations.
 Therefore, the theoretical treatment of the adsorption of polymers
 and proteins in the framework of a minimalistic coarse-grained protein 
models in statistical mechanism is a fascinating field that still attract
 interest~\cite{Eisen, Fleer, Metzger}. Coarse-grained models for polymers 
or proteins include two  monomer types which stand for  at the primitive
 order the amino acids~\cite{irbaeck, pande, vilgis}. These monomers
 classified  either hydrophobic and polar types which these models  are 
known as hydrophobic-polar (HP) or AB models. The presence of an flat 
surface strongly affects the polymers or proteins conformational 
behaviour in the near of the interface~\cite{Moddel}. Because in their
 own right the monomer-monomer attraction is responsible for  the collapse
 and the explicit surface, that is  surface-monomer attraction  will 
compete with each other. This competition will  results in folding vs.
 adsorption for polymers near an attractive surface.
 
In this letter, a minimal theoretical model  that presents some base 
trends  of polymer adsorption is studied. In order to analyze  thermodynamic properties 
and some structural parameters  of the adsorption of polymer chains near 
an unstructured, attractive flat surface  Multicanonical Monte Carlo 
simulations are performed. Firstly, the representation
 proceed by replacing the explicit attractive surface with an implicit
 potential, Lennard-Jones  potential between the monomers and the surface.
 The objective of this work is to show that with a simple, minimal model,
 it is prospective to  capture folding vs. adsorption mechanism for the
 polymer-surface system. 

This paper is organized as follows. In Sec.II the minimalistic 
coarse-grained polymer-surface  model  and the multicanonical simulation 
method are given. In Sec.III, the results are presented  and the paper is 
concluded in Sec. IV. 

\section{The Model}

The polymer chain are described by a coarse-grained hydrophobic-polar model.
A manifest off-lattice generalisation of the hydrophobic-polar (HP) model~\cite{dill1}  is the AB model~\cite{Stillinger}, where the hydrophobic monomers are
labelled by $A$ and the polar or hydrophilic ones by $B$. The contact interaction is replaced by
a distance-dependent Lennard-Jones type of potential accounting for
short-range excluded volume repulsion and long-range interaction, the latter being attractive for $AA$ and
$BB$ pairs and repulsive for $AB$ pairs of monomers. An additional interaction accounts for the bending
energy of any pair of successive bonds. This model was first applied in two dimensions~\cite{Stillinger} and
generalized to three-dimensional AB proteins~\cite{irb1,irb2},
partially with modifications taking implicitly into account additional
torsional energy contributions of each bond.

AB model as proposed in Ref.~\cite{Stillinger} with the energy function
\begin{eqnarray}
\label{stillinger}
E_{AB} &=& \frac{1}{4}\sum\limits_{k=1}^{N-2}(1-\cos \vartheta_k)+\nonumber\\
&&\hspace{7mm} 4\sum\limits_{i=1}^{N-2}\sum\limits_{j=i+2}^N\left(\frac{1}{r_{ij}^{12}}
-\frac{C_{AB}(\sigma_i,\sigma_j)}{r_{ij}^6} \right),
\end{eqnarray}
where the first sum runs over the $(N-2)$ angles $0\le \vartheta_k\le \pi$ of successive bond vectors.
This term is the bending energy and the coupling is ``ferromagnetic'', i.e., it costs
energy to bend the chain. The second term partially competes with the bending barrier by a
potential of Lennard-Jones type depending on the distance between monomers being non-adjacent
along the chain. It also accounts for the influence of the AB sequence ($\sigma_i=A$ for hydrophobic
and $\sigma_i=B$ for hydrophilic monomers) on the energy of a conformation
as its long-range behavior is attractive for pairs of like monomers and repulsive for $AB$ pairs
of monomers:
\begin{equation}
\label{stillC}
C_{AB}(\sigma_i,\sigma_j)=\left\{\begin{array}{cl}
+1, & \hspace{7mm} \sigma_i,\sigma_j=A,\\
+1/2, & \hspace{7mm} \sigma_i,\sigma_j=B,\\
-1/2,  & \hspace{7mm} \sigma_i\neq \sigma_j.\\
\end{array} \right.
\end{equation}

\begin{figure}
\psfig{figure=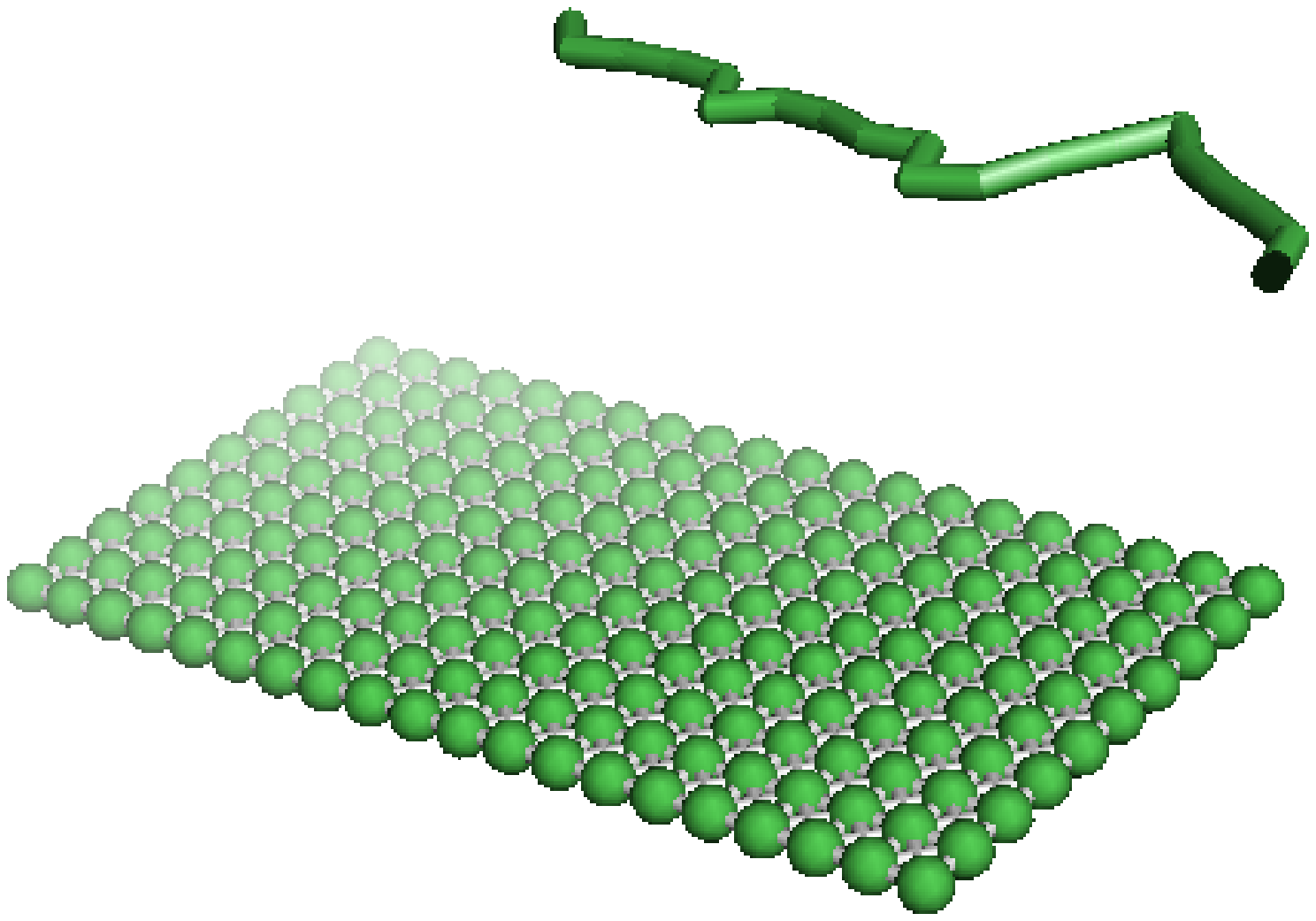,width=15cm,angle=0}
\caption{The starting configuration of the simulation. The polymer chain is an randomly start configuration and the surface  is fixed in the whole simulation. }
\label{figure1}
\end{figure}

The adsorption of a hydrophobic polymer chain near a attractive  surface  is  simulated using off-lattice AB model. It is considered that a polymer chain is over the surface enough to not to  cause reducing  the entropic freedom of the polymer. The surface  is composed of purely same  type of sites that hydrophobic monomers in the polymer chain has an attractive interaction with the surface and additionally with a surface attraction strength parameter ($\epsilon$ ) the dosage of the attraction is varied. A start configuration to the simulation is presented in  
Figure~\ref{figure1}. 

The energy function of the polymer chain is introduced above and the interaction of polymer chain monomers (m)  and surface (s)  is given with Lennard-Jones  type potential:
\begin{equation}
\label{adsorption}
E_{ms} =  4  ~ \epsilon \sum\limits_{i=1}^{N_m}\sum\limits_{j=1}^{N_s}
\left(\frac{1}{r_{ij}^{12}}
-\frac{C_{ms}(\sigma_i,\sigma_j)}{r_{ij}^6} \right),
\end{equation}
where the $N_m$ is the number of monomers and $N_s$ is the number of interaction sites in the surface. The  $ C_{ms} $ parameter is settled for one between the $AA$ contacts and  $\epsilon$ is the surface attraction strength which is varied in the simulation. Then the total energy ( $E_T $ ) of the system will contain pure AB model polymer chain energy and the polymer chain  surface interaction energy ( $ E_T = E_{AB} + E_{ms} $ ). The initial configuration of the polymer chain is randomly generated where the ends have no contact with the  surface. In some theoretical and computational studies the polymer is attached at the surface with one of its ends but this  reduces the entropic freedom of the system~\cite{Bachmann2}. However, in many recent experiments of the peptide-metal or peptide semiconductor interfaces the setup is considered by a freely moving polymer in a surface. This enables to polymer conformations to choose folding vs. adsorption behaviours without constraint.

\bigskip

\bigskip

Simulations of this model were performed with  Multicanonical
Algorithm~\cite{MUCA} which the details are given below. The update mechanism for a polymer chain is spherical update
which is described in Ref.~\cite{Arkin} in detail. The surface position is fixed in the whole simulation.
 The atomistic detailed plot of the surface is only chosen for better visualization. 

\section{Simulation Method}

The multicanonical ensemble  is based on a probability
function in which
the different energies are equally probable.
However, implementation of  the multicanonical algorithm (MUCA)
is not straightforward
because the density of states  $n(E)$ is {\it a priori} unknown.
In practice, one only needs  to know the weights  $\omega$,
\begin{equation}
 w(E) \sim 1/n(E) = \exp [(E-F_{T(E)})/k_BT(E)],
\end{equation}

and these weights are calculated in the
first stage of simulation  process  by an iterative  procedure
in which
the temperatures $T(E)$ are built recursively together with
the microcanonical free energies $F_{T(E)}/k_BT(E)$  up to an
additive constant.
The iterative  procedure is
followed by a long production run based on the fixed $w$'s where
equilibrium configurations are sampled. Re-weighting techniques~\cite{FeSw88}
enable one to obtain Boltzmann averages of various
physical variables  over a wide range of temperatures.

\bigskip

As pointed out above, calculation of the {\it a priori} unknown MUCA weights
is not trivial, requiring an experienced  intervention.
For lattice models, this problem was addressed in a
sketchy way by Berg and \c{C}elik~\cite{MUCA} and later by
Berg~\cite{Be98}.
An alternative way is to establish an automatic process by
incorporating the statistical errors within the recursion
procedure.

\section{Results and Discussions}

For each $\epsilon$ values ( $\epsilon= 0.01, 0.5, 1.0 $ ), after calculating the multicanonical weights $ 5\times 10^{7} $ iteration were performed in production run.  In literature, many recent papers~\cite{PRL97, PRE73, PRE72, PRE68, Arkin2} give only the minimum energy values for different sequences in the frame of this model, but no other aspects of physics are  investigated for this effective protein models. The main aim of  this study is  that by employing a simple coarse-grained hydrophobic-polar model and  adding an interaction term to the energy which describes the interaction  with the environment, it is indeed possible to identify  
fundamental characteristics of the physics of protein adsorption. This work can be extend or give the basis of the adsorption  process of real proteins.

\begin{figure}
\centerline{
\psfig{figure=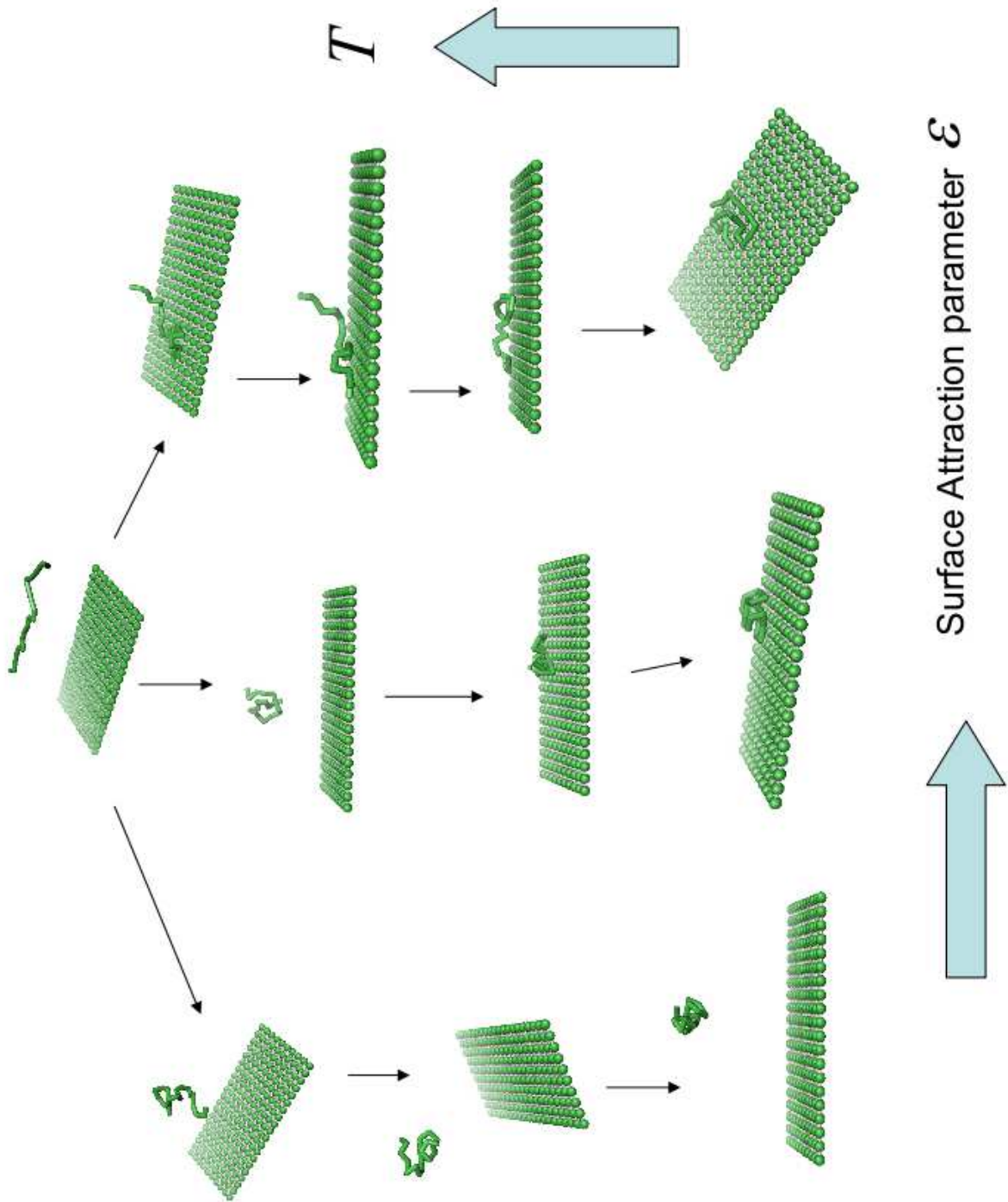,width=17cm,angle=0}} 
\caption{The chart of typical conformations for  different regions of temparature and surface attraction strength.  }
\label{figure2}
\end{figure}

In Fig~\ref{figure2},  a chart of folding vs. adsorption mechanisms are presented visually for the three $\epsilon$ values. At the top of
the chart, there is the start conformation of the all different simulations. Afterward the chart goes to
three different paths which corresponds to different surface attraction strenghts $\epsilon$. $\epsilon$
parameter and temperature increases from right to left and from bottom to top, respectively. The right column corresponds to $\epsilon= 0.01$, the middle corresponds to $\epsilon= 0.5$ and the left one corresponds to $\epsilon= 1.0$. Beginning with the right column, it is easily verified that when the temperature is decreasing  the polymers goes  from random coil conformation to collapse one and at the end it is folded to its native structure. There is no contact with the surface in other words no adsorbed conformations can be found. In the middle column, the chain is going very close to the surface
but at the same time it is also going to collapse.  The existence  of a surface strongly affects the polymers  conformational behaviour in the interface. Because in their own right the monomer-monomer attraction is responsible for  the collapse and the explicit surface, that is  surface-monomer attraction  will compete with each other. This competition gives rise to different conformational 
phases. At somewhat high temperatures, two way can be  seen in dependence of the competition between monomer-monomer and monomer-surface interactions. In first case, desorbed globule conformation  will adsorbed to the surface, in the other case absorbed but random coil state will rearranged and get its adsorbed compact conformation. Finally, if the $\epsilon$ is sufficiently greater ( the left column) then the polymer first adsorb to the surface with an expanded conformation and finally to getting a flat  two-dimensional conformation. 

\begin{figure}
\centerline{
\psfig{figure=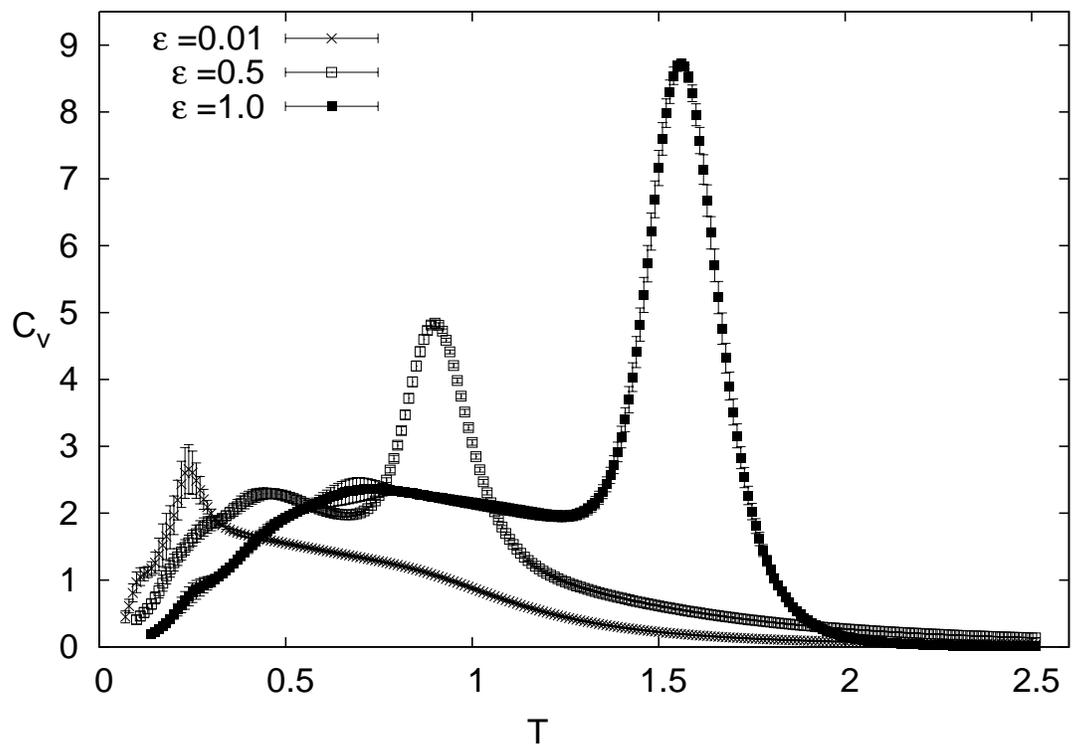,width=15cm,angle=0}}
\caption{ The specific heat as a function of temperature for different values of $\epsilon$ parameter.  }
\label{figure3}
\end{figure}

\begin{figure}
\centerline{
\psfig{figure=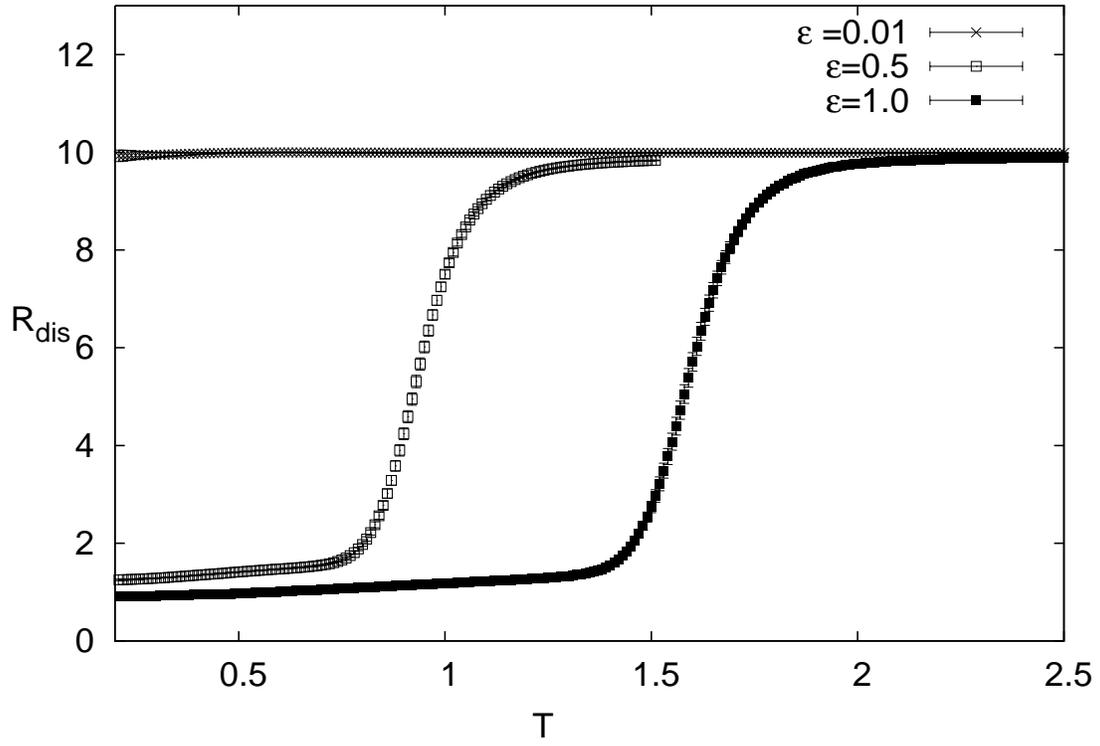,width=15cm,angle=0}}
\caption{ The average of the center of mass distance to the surface as a function of temparature for different values of $\epsilon$ parameter. }
\label{figure4}
\end{figure}

In order to get an insight about conformational transitions, the specific heats 
$C_V(T)=(\langle E^2\rangle-\langle E\rangle^2)/k_BT^2$ with
$\langle E^k\rangle=\sum_E g(E) E^k \exp(-E/k_BT)/\sum_E g(E) \exp(-E/k_BT)$ are calculated 
 by reweighting the multicanonical
energy distribution obtained with multicanonical sampling to the canonical distribution. Details are given in Ref.~\cite{Arkin}.
In Fig~\ref{figure3}  the specific heats   for the three $\epsilon = 0.01, 0.5, 1.0 $ values are given.  Firstly interpreting the curves for the specific heats in Fig~\ref{figure3} in terms of conformational transitions, it can be concluded that the specific heat curve for the $\epsilon = 0.01 $ exhibits a pronounced peak at $ T = 0.25$ which interpreted as the folding temperature.
The polymer chain  tend to form more compact conformations in this  temperature region.
 It is widely believed and experimentally consolidated that realistic short single domain proteins are usually two-state folders~\cite{chan1}. This means,
there is only one folding transition and the protein is either in the folded or an unfolded (or denatured) state. This results is consistent with our previous work~\cite{Arkin}.
But the specific heat curves for $\epsilon =  0.5, 1.0 $
exhibits one pronounced peak at even higher temperature values and one shoulder indicating two transitions in the profile of the specific heat. First one is the adsorption transition separating the desorbed and adsorbed conformations  at high temperatures and the second one is same kind of freezing 
transition which the conformations rearranged and get its compact conformations~\cite{Moddel, Arkin3}. 
An explicit demostration that the polymer is freely moving and going self-folding or is very close to the surface  can be elucidate with the distance of the center of mass of the polymer to the flat surface. The average of the center of mass distance to the surface ( $ R_{dis}$ ) are calculated and presented for the three $\epsilon$  values in  Fig~\ref{figure4}. This parameter detects nicely the distinction  between adsorption transition and self-folding of the polymer chain near the flat surfaces. As can be seen in  Fig~\ref{figure4}, for small value of $\epsilon$, the polymer chain moves freely  above the surface and  will go to self-folding . Therefore the average value of the $ R_{dis}$ is approximately constant at all temperature range. On the other hand, at higher $\epsilon$ values the  $ R_{dis}$ parameter presents a prominent turning points. For $\epsilon = 0.5$  this turning point occurs at 
$T = 0.85 $ and for $\epsilon = 1.0$  at $ T= 1.56 $ which can also detectable from the specific heat peaks in Fig~\ref{figure3}.

Finally, in order to check the structural  compactness of conformations for the folded case  or to identify the possible dispertion of conformations because of adsorption, the radius of gyration and the end-to-end distance parameters of the global-minimum conformations for different $\epsilon$ values  are calculated and given in Table~\ref{table1}. For the folded conformation which is the $\epsilon = 0.01$ case the radius of gyration and the end-to-end distance parameters are smallest. For higher values of $\epsilon$ parameters, the end-to-end distances and radius of gyrations  are increased  because of adsorption to the flat surface. 

\section{Conclusion}
In this paper, the adsorption of proteins and peptides within the framework of minimalist effective coarse-grained polymer  model is presented. As the adsorption potential, Lennard-Jones type potential, between the effective monomers and the flat  surface is assumed. By changing the attraction strenght of the flat surface folding vs. adsorption transition are detected and the typical conformations are presented.  Despite the simplicity of the model, it is possible to see some basic characteristics of the protein adsorption. The work considered in this paper will promote  practical implications for a wide sort of problems ranging from protein-ligand binding to designing smart sensors.

\pagebreak
\newpage

\pagebreak
\newpage

\pagebreak
\newpage

\pagebreak
\newpage

\pagebreak
\newpage

\begin{table}
\caption{The radius of gyration, the end-to-end distance and the average of the center of mass distance  parameters for different surface attraction strength values for the global minimum 
energy conformations. }
\label{table1}
\begin{tabular}{|c|c|c|c|}
\hline ~~~
$\varepsilon$~~~ & ~~~ R$_{gy}$~~~ & ~~~D$_{ee}$ ~~~ & ~~~ R$_{dis}$~~~\tabularnewline
\hline
\hline
0.01 & 1.23 & 1.55 & 6.61\tabularnewline
\hline
0.5 & 1.37 & 2.50 & 1.21\tabularnewline
\hline
1.0 & 2.02 & 2.66 & 1.02\tabularnewline
\hline
\end{tabular}
\end{table}


\small

\begin{thebibliography}{12}

\bibitem{Steiner} S. Walheim, E. Schaffer, J. Mlynek and U. Steiner, Science 283, 520 (1999).

\bibitem{nakata}
E.\ Nakata, T.\ Nagase, S.\ Shinkai, and I.\ Hamachi, J.\ Am.\ Chem.\ Soc.\ {\bf 126}, 
490 (2004).

\bibitem{bogner}
T.\ Bogner, A.\ Degenhard, and F.\ Schmid, Phys.\ Rev.\ Lett.\ {\bf 93}, 268108 (2004)


\bibitem{Service} R. F. Service, Science 270, 230 (1995).

\bibitem{balog}
E.\ Balog, T.\ Becker, M.\ Oettl, R.\ Lechner, R.\ Daniel, J.\ Finney, and J.\ C.\ Smith,
Phys.\ Rev.\ Lett.\ {\bf 93}, 028103 (2004). 

\bibitem{ike}
M.\ Ikeguchi, J.\ Ueno, M.\ Sato, and A.\ Kidera, Phys.\ Rev.\ Lett.\ {\bf 94}, 078102 (2005).

\bibitem{gupta}
N.\ Gupta and A.\ Irb\"ack, J.\ Chem.\ Phys.\ {\bf 120}, 3983 (2004).


\bibitem{Bachmann1} M. Bachmann and W. Janke, Phys. Rev. Lett. 95, 058102 (2005).

\bibitem{Bachmann2} M. Bachmann and W. Janke, Phys. Rev. E 73, 041802 (2006).

\bibitem{Gokoglu} G. G\"{o}ko\u{g}lu, M. Bachmann, T. \c{C}elik, W. Janke, Phys. Rev. E 74, 041802 (2006).


\bibitem{PRE73R}   M. Bachmann and W. Janke, Phys. Rev. E 73, 020901(R) (2006).


\bibitem{Eisen} E. Eisenriegler, "Polymers near Surfaces", World Scientific, Singapore 1993.

\bibitem{Fleer} G. J. Fleer, M. A. Cohen-Stuart, J. M. H. M. Scheutjens, T. Cosgrove, B. Vincent, Polymers at Interfaces, Chapman and  Hall, London 1993.

\bibitem{Metzger} S. Metzger, M. Müller, K. Binder, J. Baschnagel, Macromol. Theory Simul. 11, 985 (2002).


\bibitem{irbaeck} A.\ Irb\"ack, C.\ Peterson, F.\ Potthast, Proc. Natl. Acad. Sci. U.S.A. 93, 9533 (1996).

\bibitem{pande} V. Pande, A. Grosberg and T. Tanaka, Biophys. J. 73, 3192 (1997).

\bibitem{vilgis} N. -K. Lee and T. A. Vilgis, Phys. Rev. E 67, 050901(R) (2003).

\bibitem{Moddel} M. M\"{o}ddel, M. Bachmann, W. Janke, J. Phys. Chem. B 113, 3314 (2009). 

\bibitem{dill1}
K.\ A.\ Dill, Biochemistry {\bf 24}, 1501 (1985); K.\ F.\ Lau and K.\ A.\ Dill,
Macromolecules {\bf 22}, 3986 (1989).


\bibitem{Stillinger}
F.\ H.\ Stillinger, T.\ Head-Gordon, and C.\ L.\ Hirshfeld, Phys.\ Rev.\ E {\bf 48}, 1469 (1993);
F.\ H.\ Stillinger and T.\ Head-Gordon, Phys.\ Rev.\ E {\bf 52}, 2872 (1995).


\bibitem{irb1}
A.\ Irb\"ack, C.\ Peterson, F.\ Potthast, and O.\ Sommelius, J.\ Chem.\ Phys.\ {\bf 107}, 273 (1997).

\bibitem{irb2}
A.\ Irb\"ack, C.\ Peterson, and F.\ Potthast, Phys.\ Rev.\ E {\bf 55}, 860 (1997).

\bibitem{MUCA} B.A. Berg, T. \c{C}elik,
Phys Rev Lett 69, 2292 (1992);  B.A. Berg, Fields Institute Communications 28,
1 (2000).


\bibitem{Arkin}
M.\ Bachmann, H.\ Ark{\i}n, and W. Janke, Phys.\ Rev.\ E {\bf 71}, 031906 (2005).


\bibitem{FeSw88}  A.M. Ferrenberg, R.H. Swendsen, Phys Rev Lett
 61, 2635 (1988); ibid 63, 1658 (1989).



\bibitem{Be98} B.A. Berg, Nucl Phys B (Proc. Suppl.) 63A-C,
982 (1998).


\bibitem{PRL97} J. Kim, J. E. Straub, T. Keyes, Phys. Rev. Lett. 97, 050601 (2006).

\bibitem{PRE73} V. Elser, I. rankenburg, Phys. Rev. E 73, 026702 (2006).

\bibitem{PRE72} S.-Y. Kim, S. B. Lee, J. Lee, Phys. Rev. E, 011916 (2005).

\bibitem{PRE68} H.-P. Hsu, V. Mehra, P. Grassberger, Phys. Rev. E 68, 037703 (2003). 

\bibitem{Arkin2}
 H.\ Ark{\i}n, Phys.\ Rev.\ E {\bf 78}, 041914 (2008).

\bibitem{chan1}
H.\ S.\ Chan, S.\ Shimizu, and H.\ Kaya, Meth.\ in Enzym.\ {\bf 380}, 350 (2004).

\bibitem{Arkin3} H. Arkin,  Phys.\ Rev.\ E {\bf 80}, 041910 (2009).


\end{thebibliography}
\end{document}